\documentclass[a4paper,12pt]{article}
\usepackage[cp1251]{inputenc}
\usepackage[dvips,dvipdf]{graphicx,graphics}
\usepackage[english]{babel}
\usepackage{amsfonts,amssymb, amsmath, amsthm}

\title{State equation for the three-dimentional system of \textquotedblleft collapsing\textquotedblright\ hard spheres\footnote{This article is the revised and superset version of our work arXiv:cond-mat/0604239}}
\date{}
\author{I.\,Klebanov**, N.\,Ginchitskii**, P.\,Gritsay**}

\begin{document}
\maketitle

\begin{quotation}
\footnotesize
\begin{center}
\itshape
**The Chelyabinsk State Pedagogical University, the Department of Mathematics\\
\end{center}

By Wertheim method the exact solution of the Percus-Yevick integral equation for a system of particles with the \textquotedblleft repulsive step potential\textquotedblright\,interacting (\textquotedblleft collapsing\textquotedblright\,hard spheres) is obtained. On the basis of this solution the state equation for the \textquotedblleft repulsive step potential\textquotedblright\,is built and determined, that the Percus-Yevick equation does not show the Van der Waalse loop for \textquotedblleft collapsing\textquotedblright\,hard spheres.

\end{quotation}

During the last 15 years the system of \textquotedblleft collapsing\textquotedblright\, hard spheres (CHS) has been of the greatest interest of some experts in the field of physics of the condensed matter. The system of CHS is the system of particles with pair interaction potential
\begin{equation}
V(r)=
	\begin{cases}
		\infty, &r<R\\
		V_{0},&R\leq r\leq R+a\\
		0,&r>R+a,     
	\end{cases}
\end{equation}
where $V_{0}$ is the positive constant, $r$ is the particles range, $R$ is the hard core diameter, $a$ is the soft core diameter [see \cite{Santos:06} and the references given there].This interest is connected with a great application of the system of CHS in mathematical modeling isomorphic phase transitions, transformations in colloid systems, in the study of curve of fusion, etc. \cite{Ryzhov:03}. By this time the calculations of thermodynamic characteristics of the system of CHS have been made by means of molecular dynamics and thermodynamic perturbation theory within the framework of spin-liquid approach [see for example \cite{Mikheenkov:04} and the references given there].\\
However, there is another approach to the study of system with interaction in statistical mechanics, i.e. the solution of approximate integral equations for the pair correlation function. We have not come across the case of solutions of integral equation for the pair distribution function of the system of CHS. In our opinion this research is necessary for the study of possibilities of integral equations method, because the use of this method in liquid theory for describing phase transitions is still the subject of many discussions \cite{Martynov:99}. 
In this work the system of CHS is being studied in the approach of Percus-Yevick (PY). It is an open secret that the PY equation admits the exact analytical solution in case of "hard spheres" potential \cite{Wertheim:63}-\cite{Santos:99}. Up to now it is the only analytical solution of nonlinear integral equation for the pair distribution function. On the basis of the method suggested in \cite{Wertheim:64} we have obtained the exact analytical solution for a more complicated and "realistic" system of CHS for three-dimensional space when the number of particles and temperature are constant. 
Let's consider the PY equation in three diemensions (we follow the signs used in \cite{Wertheim:64})
\begin{equation}
\tau(\textbf{r})=1-\rho\int\tau(\textbf{r})f(\textbf{r})\,d\textbf{r}+\rho\int\tau(\textbf{r}^{\prime})f(\textbf{r}^{\prime})\tau(\textbf{r}-\textbf{r}^{\prime})e(\textbf{r}-\textbf{r}^{\prime})\,d\textbf{r}^{\prime},
\end{equation}
where $e(x)\equiv e^{-\beta V(x)}, f(x)=e(x)-1,\, \rho$ is the density, and $\beta=(kT)^{-1}$. The pair distribution function $g(x)$ and the direct correlation function $C(x)$ of Ornstein and Zernike are in the PY approximation related to $\tau(x)$ by
\begin{equation}
g(x)=\tau(x)e(x),\quad C(x)=\tau(x)f(x).
\end{equation}
We make the equations dimensionless by defining $x=\left(r/R\right)$, $\eta=\left(\pi R^{3}\rho/6\right)$, $a^{\prime}=a/R$. Then we move to bipolar coordinates and integrate for angle variable for the repulsive step potential, and take the one side Laplace transform of the equation satisfied by $h(x)\equiv x \tau(x)$.
Defining
\begin{equation}
\begin{split}
&F(t)\equiv-\int\limits_0^{1+a^{\prime}}xC(x)e^{-tx}dx=\int\limits_0^1 h(x)e^{-tx}dx+\epsilon_2\int\limits_1^{1+a^{\prime}}h(x)e^{-tx}dx,\\
&G(t)\equiv \int\limits_1^\infty xg(x)e^{-tx}dx=\epsilon_1\int\limits_1^{1+a^{\prime}}h(x)e^{-tx}dx+\int\limits_{1+a^{\prime}}^\infty h(x)e^{-tx}dx,\\
&K\equiv -F^{\prime}(0)=-\int\limits_0^{1+a^{\prime}}x^2 C(x)dx=\int\limits_0^1 xh(x)dx+\epsilon_2\int\limits_1^{1+a^{\prime}}xh(x)dx,\\
&\Upsilon(t)\equiv \int\limits_0^{a^{\prime}}y(x)e^{-tx}dx,\\
&y(x)\equiv -\int\limits_{1+x}^{1+a^{\prime}}x^{\prime}C(x^{\prime})(x-x^{\prime})g(x-x^{\prime})dx^{\prime}=\epsilon_1\epsilon_2\int\limits_{1+x}^{1+a^{\prime}}h(x^{\prime})h(x-x^{\prime})dx^{\prime}\\
&\epsilon_1=e^{-\beta V_0},\quad \epsilon_2=1-\epsilon_1\\ 
\end{split}
\end{equation}
we obtain
\begin{equation}
t \left[ F(t)+G(t)\right]=\frac{1+24\eta K}{t}+12\eta \left\{\left[F(t)-F(-t)\right]G(t)-\Upsilon(t)+\Upsilon(-t)\right\}.
\end{equation}
Solving for G(t), this becomes
\begin{equation}
G(t)=\frac{\cfrac{1+24\eta K}{t^{2}}-F(t)-\cfrac{12\eta}{t}\left[\Upsilon(-t)-\Upsilon(t)\right]}{1+\cfrac{12\eta}{t}\left[F(-t)-F(t)\right]}.
\end{equation}
For further investigation we define the folowing function
\begin{equation}
H(t)=t^4 G(t)\left\{\frac{1+24\eta K}{t^2}-F(-t)+\frac{12\eta}{t}\left[\Upsilon(-t)-\Upsilon(t)\right]\right\}+\Upsilon(t)+\Upsilon(-t).
\end{equation}
Discussion like in \cite{Wertheim:64} show that
\begin{equation*}
H(t)=\lambda_1+\lambda_2t^2+\lambda_3t^3,
\end{equation*}
where $\lambda_1, \lambda_2$ are constants. However, it is easy to check that $H(t)$ is an even function, and so we define
\begin{equation*}
H(t)=\lambda_1+\lambda_2t^2.
\end{equation*}
Not including $G(t)F(-t)$ from (6) and (7) and turning the Laplace transform into the $x<1+a^{\prime}$, we gain the explicit expression for $h(x)$
\begin{equation}
h(x)=C_0+C_{1}x+C_{2}x^2+C_{4}x^4,
\end{equation}
where $C_i$ is the solution of the following system of equations  
\begin{equation}
\begin{cases}
\begin{split}
&C_0=b\left[\epsilon_1 f_{0}^{(2)}\left(f_{0}^{(1)}+\epsilon_2f_{0}^{(2)}\right)-\frac{1}{12}y_4-\alpha\epsilon_1
f_{2}^{(2)}-2b\epsilon_1 y_1 f_{0}^{(2)}\right]\\
&C_1=\alpha\left[1+b\epsilon_1 f_{1}^{(2)}\right]\\
&C_2=-\frac 1 6 b\left[bf_{3}^{(1)}+b\epsilon_2 f_{3}^{(2)}-12by_1+6f_{0}^{(1)}+3f_{0}^{(2)}\left(2\epsilon_2+
\alpha\epsilon_1\right)\right]\\
&C_4=\frac{1}{24}\alpha b
\end{split}
\end{cases}
\end{equation}
and $\epsilon_1=e^{-\beta V_0}, \epsilon_2=1-\epsilon_1, b=12\eta$.
The constants $\alpha, f_{n}^{(1)}, f_{n}^{(2)}$ and $y_n$ as density, temperature and potential parametres $V_0, R, R+a$ function can be obtained from
\begin{equation}
\begin{split}
&f_{n}^{(1)}=\displaystyle\int\limits_0^1 x^n h(x)dx\\
&f_{n}^{(2)}=\displaystyle\int\limits_1^{1+a^{\prime}} x^n h(x)dx\\
&\alpha=1+24\eta \left(f_{1}^{(1)}+\epsilon_2 f_{1}^{(2)}\right)\\
&y_{n}=\displaystyle\int\limits_0^{a^{\prime}}x^n y(x)dx\\
&y(x)=\epsilon_1 \epsilon_2\displaystyle\int\limits_{1+x}^{1+a^{\prime}}h(x^{\prime})h(x-x^{\prime})dx^{\prime}.
\end{split}
\end{equation}
The system (9) must be completed with the condition $C_0=0$, because $h(0)=0$ as it is followed from (2) and the nature of $h(x)$.
The system of \textquotedblleft collapsing\textquotedblright\, hard spheres state equation can be shown as
\begin{equation}
\begin{split}
\frac{P}{\rho kT}&=1-\frac{\rho}{6kT}\int \tau(\textbf{r})\frac{dV}{d\textbf{r}}d\textbf{r}=\\
&=\frac{16\eta T}{\pi}\left(1+4\eta\left[\epsilon_1 h(1)+\epsilon_2\left(1+a^{\prime}\right)^2 h(1+a^{\prime})\right]\right)
\end{split}
\end{equation}
and inverse isothermic compressebility
\begin{equation}
\left(\frac{\partial P}{\partial \rho}\right)_{T}\frac{1}{kT}=1-\rho\int C(\textbf{r})d\textbf{r}=\lambda_{1}(\rho,T).
\end{equation}
It can be seen from (12) and (10) that if $V_0>0$
\begin{equation}
\left(\frac{\partial P}{\partial\rho}\right)_T>0
\end{equation}
i.e. the Van der Waalse loop is absent in the isotherm.\\
The gained solutions admits the generalization in case of n-step potential of particles. The calculations show that in case of $n\geq 2$ the Van der Waalse loop exists. Besides, the developed method allows to gain approximate solutions of the Percus-Yevick equation and, therefore, state equations for any smooth potential of particles. For this reason it is enough to appriximate the smooth potential with the n-step one. The systems of two-step potential are of the major interest nowadays \cite{Skibinsky:03}-\cite{Franzese:00}. The state equation of such systems in the Percus-Yevick approximation is the issue of our next article.

\end{document}